\documentstyle[11pt]{article}
\def\ave#1{\left\langle #1\right\rangle}

\font\rm=cmr10 scaled\magstep1

\def\text{\rm}

\newcommand{\Ket}[1]{|#1\rangle}
\newcommand{\Braket}[2]{\langle #1|#2\rangle}

\begin{document}
\begin{flushright}
\today
\end{flushright}
\begin{center}
\vspace{0.3in}
{\Large\bf  Quantization of generic chaotic 3D billiard with smooth 
boundary II: structure of high-lying eigenstates}\\
\vspace{0.4in}
\large
Toma\v z Prosen
\footnote{e-mail: prosen@fiz.uni-lj.si}\\
\normalsize
\vspace{0.3in}
Physics Department, Faculty of Mathematics and Physics,\\
University of Ljubljana, Jadranska 19, 1000 Ljubljana, Slovenia\\
\vspace{0.3in}
\end{center}
\vspace{0.5in}

\noindent{\bf Abstract}
This is the first survey of highly excited eigenstates of a chaotic 3D billiard.
We introduce a strongly chaotic 3D billiard with a smooth boundary and we manage to
calculate accurate eigenstates with sequential number (of a 
48-fold desymmetrized billiard) about 45,000. 
Besides the brute-force calculation of 3D wavefunctions we propose and illustrate 
another two representations of eigenstates of quantum 3D billiards: 
(i)  normal derivative of a wavefunction over the boundary surface, and 
(ii)  ray --- angular momentum representation.
The majority of eigenstates is found to be more or less uniformly extended over the
entire energy surface, as expected, but there is also a fraction of strongly 
localized --- scarred eigenstates which are localized either (i) on to classical periodic
orbits or (ii) on to planes which carry (2+2)-dim classically invariant manifolds,
although the classical dynamics is strongly chaotic and non-diffusive.

\vspace{0.3in}



\newpage

\noindent
The structure of individual quantum eigenstates of classically chaotic closed systems
is one of the most important and difficult questions of quantum chaos.
Understanding of  localization properties of highly excited eigenstates can explain
almost any statistical property of a quantum system, such as e.g. statistics of energy
levels (see e.g. \cite{BTUPR}) or statistical distribution of transition amplitudes
\cite{P94}.
For classically fully chaotic --- ergodic billiards we have a theorem
\cite{SCV} which states that asymptotically, in the semiclassical limit
$\hbar\rightarrow 0$,  probability density of eigenstates should become uniformly
delocalized --- extended over the billiard domain, i.e. phase space distributions of
eigenstates should approach a uniform distribution over the energy surface.
The fluctuations of an eigenfuction $\Psi(\vec{r})$ are also expected to obey
a universal law for classically fully chaotic systems \cite{B77}. For systems having
a time-reversal symmetry, $\Psi(\vec{r})$ is expected to behave like a real
Gaussian random variable $\Psi$ with a uniform probability distribution
$d{\cal P}/d\Psi = 1/(\sqrt{2\pi}\sigma)\exp(-\Psi^2/2\sigma^2)$
where inverse variance $1/\sigma^2$ is constant and equal to the volume of the billiard.

For finite values of Planck's constant $\hbar$ there may be considerable deviations 
from this semiclassical
limit: (i) There may be classical adiabatic invariants, which change
slowly and diffusively in the course of time, or partial barriers in phase space ---
cantori, so that classical orbits spend a long time to uniformly 
fill entire energy surface.
If this time is longer than Heisenberg ({\em break}) time
$t_{\text break} = \hbar d(E)$, where $d(E)$ is the density of states,  then one should
expect strong localization of quantum eigenstates on such approximately
disconnected components of energy surface \cite{BTUPR}. 
(ii) But even if classical dynamics
quickly explores the entire energy surface w.r.t. $t_{\text break}$ 
one can get strong enhancements of
probability density of eigenstates on the least unstable classical periodic
orbits --- the so-called {\em scars} \cite{H84}.

Semiclassical theory has been developed, 
which explains the effect of classical periodic orbits
onto the composition of eigenstates within an energy interval much larger than the mean
level spacing
\cite{BBAF}, but there is no adequate theory to predict the structure of 
individual eigenstates.
One can use a heuristic argument due to Heller \cite{H84}, saying that an eigenstate
is scarred by a given classical periodic orbit if a Gaussian wavepacket,
which is launched along the orbit, interferes constructively
with itself after one period. But in chaotic systems there are many
periodic orbits and it is typically impossible to tell afortiori
which orbits will scar most intensively. (The rule that these are the
least unstable ones holds only on average.)

In the literature there exists a vast amount of numerical
evidence on scarring in chaotic systems with two freedoms
(see e.g. \cite{PR93},\cite{LR94},\cite{AOCO}), 
while the structure of comparably high-lying eigenstates of three and
higher dimensional chaotic systems have numerically not yet been studied,
so it is not known whether classical periodic orbits can scar
eigenstates in higher dimensions or not. Heuristic geometrical argument
might suggest that scarring by individual periodic orbit is less
effective in 3D than in 2D since more phase space is available for a
wavepacket to disperse before it can interfere with itself.
But scarring by families of similar periodic orbits \cite{PR93}
may be effective in higher dimensions.
Actually, as approaching the limit $\hbar\rightarrow 0$, quantum invariant states ---
eigenstates should approach classical invariant states which are spanned by
characteristic functions on the classically invariant (sub)manifolds in phase space;
but only the {\em microcanonical} distribution is stable for long times in a fully chaotic 
system, so it is the only one which survives the semiclassical limit. 
Periodic orbits are only special cases of
classically invariant submanifolds in the phase space of a dynamical system
((1+1)D invariant submanifolds when they are parametrized by the energy $E$).
In a 3D system there may exist also (2+2)D invariant submanifolds in 6D phase space 
which can have 2D (or 3D) projection onto 3D configuration space supporting a 
2D (or 3D) scar. This type of localization is a genuine 3D effect which is yet to be 
numerically confirmed.
\\\\
The purpose of this paper is to give the first survey
of the high-lying eigenfunctions of a strongly chaotic 3D system.
We have defined a family of {\em generic} 3D billiards with a smooth
$C^\infty$ boundary (see also \cite{P96}).
Since no {\em generic} system is known to be rigorously ergodic we have
defined a two-parameter family of 3D billiards whose shapes are given by
simplest smooth deformations of
a sphere: The radial distance $r_B(\vec{n})$ from the origin to the
boundary as a function of the direction $\vec{n}, n^2=1$, is
\begin{equation} 
r_B(\vec{n}) = 1 + a (n_x^4 + n_y^4 + n_z^4) + b n_x^2 n_y^2
n_z^2, \label{eq:shape} 
\end{equation}
and contains the two lowest order terms
which preserve the cubic symmetry  
(the first two fully symmetric type ($\alpha$) cubic
harmonics after Von Lage and Bethe \cite{VLB47}).
After a careful numerical exploration of a parameter space $(a,b)$ we have
decided to chose: $a=-1/5,b=-12/5$. 
For these values of the parameters the classical
billiard is strongly chaotic: nondiffusive, without partial barriers
in phase space, and with large average maximal liapunov exponent
$\ave{\lambda_{\rm max}} = 0.54$ meaning that roughly after five
bounces we loose one digit of information of a classical orbit. There is a
tiny regular component of phase space whose relative volume has been
estimated to be $\rho_1\approx 10^{-3}$.
The billiard domain is marginally convex and there are few isolated
neutrally stable --- parabolic periodic orbits which touch the
boundary at the points of zero curvature radii.

The billiard is invariant under 48-fold cubic symmetry group $O_h$.
So we desymmetrize it: In the following we consider a 1/48 of an
original billiard (\ref{eq:shape}) with $x \ge y \ge z \ge 0$, i.e.
we consider a 3D billiard in a domain, bounded by the
boundary surface ${\cal B} = \{\vec{r}; r_B(\vec{r}/r) = r\}$, 
and the symmetry planes $z=0$, $x=y$, and $y=z$.
The symmetry planes in phase space,
$ {\cal R}_z = \{z = 0, p_z = 0\}$,
$ {\cal Q}_{xy} = \{ x = y, p_x = p_y \}$,
and
$ {\cal Q}_{yz} = \{ y = z, p_y = p_z \}$,
and the boundary surface with the tangential momentum
$ {\cal S} = \{r_B(\vec{r}/r) = r,\vec{p}\cdot\nabla
(r_B(\vec{r}/r)-r)=0\}$ are invariant (2+2)D submanifolds with respect to the
classical dynamics which have 2D projections onto 3D configuration space.
When $b=12 a$, as is the case with our choice, the billiard (\ref{eq:shape})
possesses another less trivial example of 4D invariant submanifold, namely
$ {\cal T} = \{ x = y + z, p_x = p_y + p_z\}$, with a 2D configurational
projection, the plane $x = y + z$.

Quantum eigenfunctions of our billiard (\ref{eq:shape}) carry
ireducible representations of a cubic group. We have decided to study only
the singlet eigenstates belonging to fully symmetric 1D irrep.
This corresponds to a study of a desymmetrized billiard, Helmholz
equation $(\nabla^2 + k^2)\Psi_k(\vec{r}) = 0$, with
Von Neuman boundary conditions on the symmetry planes
$z=0,x=y$ and $y=z$, and Dirichlet boundary conditions on the boundary
surface ${\cal B}$.
Quantum eigenfunctions $\Psi(\vec{r})$ have been calculated using
3D generalization of {\em scaling method}, proposed recently by 
Vergini and Saraceno \cite{VS95}.
For billiards whose shape is geometrically not very far from a sphere,
we propose to use spherical waves $\phi_{klm}(\vec{r}) =
j_l(kr)Y_{lm}(\vec{r}/r)$ insetead of plane waves $\exp(i\vec{k}\cdot\vec{r})$
as the basis of scaling functions which already solve the Helmholz equation
but do not satisfy the boundary conditions.
We define the efficiency of the
basis of spherical scaling functions \cite{P96} as $\eta = V/V_{\text sph}$
where $V$ is the volume of the billiard, and $V_{\text sph}$ is the volume of
the smallest sphere enscribed to the billiard which has radius
$r_{\text max} = \max \{ r_B(\vec{r}/r)\}$.
For a desymmetrized billiard
one should use only the linear combinations of spherical waves which already
satisfy Von Neuman boundary conditions on the symmetry planes. These
are the products of spherical Bessel functions and the
fully symmetric cubic harmonics (of type ($\alpha$) after Von Lage and Bethe
\cite{VLB47})) --- {\em cubic waves}, labeled by $\alpha$,
$$ \phi_{k\alpha}(\vec{r}) = j_{l_\alpha}(kr)
\sum_{j=0}^{l_\alpha/4} 
c_{\alpha j} (Y_{l_\alpha,4j}(\vec{r}/r) + Y_{l_\alpha,-4j}(\vec{r}/r)).$$
The coefficients $c_{\alpha j}$ are determined by the Gram-Schmidt
orthogonalization of the columns of the projector 
onto the fully symmetric irrep of the cubic group 
$O_h, P = (1/48)\sum_{G\in O_h} G$, in the basis of  
spherical harmonics $Y_{l_\alpha,4j}$ with even angular momentum $l_\alpha=2n$,
$P^{(n)}_{j j^\prime} = (1 + 2 D^{(2n)}(0,\pi/2,0)_{4j,4j'})/3$
(using the standard notation \cite{T64}).

For an accurate quantization at fixed wavenumber $k$ we 
need cubic waves $\alpha$  with angular momentum $l_\alpha$ up to 
$$ l_{\text max}(k) := k r_{\text max} + \Delta l_{\text evanescent}.$$ 
The number of included evanescent angular momenta 
$\Delta l_{\text evanescent}$ is found to be proportional to the number of accurate 
digits of the results, and being much smaller than $k r_{\text max}$.
For any {\em even} $l,\; l \le l_{\text max}$ there are 
$\lfloor l/12\rfloor + 1 - \delta_{r,2}$ fully symmetric cubic waves, where
$l = 12\lfloor l/12\rfloor + r, 0 \le r <12$, which sum up to the total number 
$$N_{\text CW}(k) = l_{\text max}(k)^2/48 + l_{\text max}(k)/4 + {\cal O}(1)$$
of fully symmetric cubic waves $\phi_{k\alpha}(\vec{r})$ which 
should accurately capture the 
eigenfunction $\Psi_k(\vec{r})$ at the wavenumber $k$
\begin{equation}
\Psi_k(\vec{r}) = \sum_\alpha \psi_\alpha \phi_{k\alpha}(\vec{r}).
\label{eq:expansion}
\end{equation}
One should find such $k$ and coefficients $\psi_\alpha$ that $\Psi(\vec{r})$  
satisfy the boundary condition on the boundary surface ,
$\Psi\vert_{\cal B} = 0$.
Following \cite{VS95} we minimize a special {\em boundary norm}
\begin{equation}
f(k) = \int\limits_{\cal B} \frac{d^2 S}{\vec{\nu}\cdot\vec{r}} |\Psi_k(\vec{r})|^2,
\label{eq:bnorm}
\end{equation}
where $d^2 S$ is a boundary surface element and 
$\vec{\nu} = \nabla(r_B(\vec{r}/r)-r)/|\ldots|$ is a unit vector
normal to the boundary ${\cal B}$,
by solving the following generalized eigenvalue problem 
\begin{equation}
\sum_{\alpha^\prime} 
\frac{d}{dk} F_{\alpha \alpha^\prime}(k_0) \psi_{\alpha^\prime} =
\lambda  \sum_{\alpha^\prime} F_{\alpha \alpha^\prime}(k_0) \psi_{\alpha^\prime},
\label{eq:gen}
\end{equation}
where 
$$ F_{\alpha\alpha^\prime}(k_0) =  \int\limits_{\cal B} \frac{d^2 S}{\vec{\nu}\cdot\vec{r}}
\phi_{k_0 \alpha}(\vec{r})\phi_{k_0 \alpha^\prime}(\vec{r}),$$
is a positive definite matrix $N_{\text CW}(k_0)\times N_{\text CW}(k_0) $.
Every matrix element is an integral over the surface of the a sphere (or over 1/48 of it),
with integration measure written in spherical coordinates $(\theta,\phi)$ 
$$
\frac{d^2 S}{\vec{\nu}\cdot\vec{r}} = d\cos\theta d\phi
\sqrt{r^2_B + (\partial_\theta r_B)^2 + (\partial_\phi r_B/\sin\theta)^2},$$
where
$ r_B = r_B(\sin\theta\cos\phi,\sin\theta\sin\phi,\cos\theta)$,
which is accurately numerically evaluated using a product grid of a Gaussian 
quadrature for $\cos\theta$ integration $\int_0^1 d\cos\theta$ 
with $l_{\text max}/2$ (positive) nodes 
and a uniform grid for $\phi$ integration $\int_0^{\pi/4}d\phi$ 
with $l_{\text max}/4$ nodes.
\footnote{
We have also implemented an optimal cubature formula due to Lebedev for accurate 
numerical integration  over the sphere, which reduces the size of a grid for integration
over the desymmetrized boundary surface ${\cal B}$, w.r.t. product grid, by a factor of  
$9/2$, and the total numerical labour by $20-40\%$.
Unfortunately we found that numerical computation
of the grid points and weights of such an optimal grid for large $l_{\text max}$ 
turns out to be extremely difficult numerically, more than the problem itself.} 
The eigenvalues $\lambda^{(n)}$ of largest modulus are good numerical estimates of 
billiard's eigenvalues $k^{(n)} \approx k_0 - 2/\lambda^{(n)}$.
Due to the scaling symmetry, 
$\phi_{k,\alpha}(\vec{r}) = \phi_{k_0,\alpha}(k\vec{r}/k_0)$,
the corresponding eigenvectors $\psi^{(n)}_\alpha$  
yield very accurate eigenfunctions 
$\Psi^{(n)}(\vec{r})$, via expansion (\ref{eq:expansion}), with error being of the order
${\cal O}((k^{(n)}-k_0)^4)$. Accurate eigenvalues have been obtained by
minimizing the boundary norm $f(k)$ (as expanded in a Taylor series around $k_0$
up to 8th order) for fixed coefficients $\psi^{(n)}_\alpha$. 
The number of converged eigenvalues $k^{(n)}$ or eigenvectors $\psi^{(n)}_\alpha$ 
in a single diagonalization of  (\ref{eq:gen}) depends on the required
accuracy,
empirically we find \cite{P96} that it is proportional to $k^2$.
\\\\
We have chosen $k_0 = 360.0$ and calculated a stretch of  $64$ consecutive eigenstates
with eigenvalues $k^{(n)}$ from a narrow window centered around $k_0$,
corresponding to sequential quantum number of a desymmetrized billiard 
${\cal N}(k_0) = 45,103 \pm 2$ (using 3D Weyl formula \cite{P96}).
Computation has been performed with $l_{\text evanescent} = 40$ 
(and $\l_{\text max}=364$) and the boundary norm (\ref{eq:bnorm}) of
normalized wavefunctions $\Psi^{(n)}$ has ranged between 
$5\cdot 10^{-9}$ and $5\cdot 10^{-8}$.
However, it is much easier to calculate high-lying eigenfunctions of a 3D billiard, than
to analyze and faithfully represent them without loosing much of information (since we
can plot functions of at most two variables). For the presentation we have picked a 
sample of 12 eigenstates out of 64: 
11 consecutive states and a specially chosen state with the criterion of being most
intensively localized.

The first representation: {\em normal derivative on the 
boundary surface}
$\vec{\nu}\cdot\nabla \Psi^{(n)}|_{\cal B}$. 
Although this representation formally contains all the information about a given 
eigenstate, it explicitly displays only a very limited portion of configuration space and
it is not very suitable to detect localization which can take place strictly 
inside the billiard region. Nevertheless, it is the easiest to be calculated numerically,
$\vec{\nu}\cdot\nabla \Psi^{(n)} = \sum_\alpha \psi^{(n)}_\alpha
\vec{\nu}\cdot\nabla \phi_{k^{(n)},\alpha},$
one only has to calculate the normal derivatives of the cubic waves on ${\cal B}$.
In figure 1 we show the nodal lines of 
$\vec{\nu}\cdot\nabla \Psi^{(n)}$ on ${\cal B}$
and the contour plots of its magnitude.
Actually, for easier planar presentation we map the desymmetrized boundary ${\cal B}$,
which has a shape of a deformed spherical triangle with angles
$\pi/4,\pi/2,\pi/3$, at the points 
$A_1=(1,0,0)r_B(1,0,0),
  A_2=2^{-1/2}(1,1,0)r_B(2^{-1/2},2^{-1/2},0),
  A_3=3^{-1/2}(1,1,1)r_B(3^{-1/2},3^{-1/2},3^{-1/2}),$
where, respectively, 4-fold, 2-fold, and 3-fold symmetry axis
penetrates ${\cal B}$, onto the half-square 
$\{(\phi,\chi),0 \le \chi \le \phi \le \pi/4\}:$
\begin{eqnarray*}
\vec{r}(\phi,\chi)  &=& \vec{n}(\phi,\chi) r_B(\vec{n}(\phi,\chi)),\\
\vec{n}(\phi,\chi)  &=& ((1-z(\chi)^2)^{-1/2}\cos\phi,
(1-z(\chi)^2)^{-1/2}\sin\phi,z(\chi)),\\
z(\chi) &=& (1 + (\sin\chi)^{-2})^{-1/2}
\end{eqnarray*}
Note that this mapping is not far from being area and angle preserving.
So we plot 
$$B^{(n)}(\phi,\chi) = \vec{\nu}\cdot\nabla\Psi^{(n)}\vert_{\vec{r}(\phi,\chi)}.$$
Figure 1 appears similar to the pictures of eigenfunctions of 2D chaotic billiards.
On diagrams we indicate the projections of known (2+2)D invariant submanifolds:
the symmetry planes are mapped on the boundary of the triangle 
${\cal R}_z \rightarrow \chi = 0,\;{\cal Q}_{xy} \rightarrow \phi = \pi/4,\;
{\cal Q}_{yz} \rightarrow \chi = \phi$,
while the invariant plane ${\cal T}$ is mapped on the curve
$\sin\chi = \cos\phi-\sin\phi$. Most of states appear more or less random
(the statistical distribution of normal derivatives over ${\cal B}$ has been analyzed
and found to be very close to a Gaussian for most of states). 
However the state
12 is strongly enhanced in the vicinity of point $A_1$, as well as the states $7,8,10$,
while the states $7,10$ are significantly enhanced in the vicinity of the projected
manifold ${\cal T}$.

Uniform representation of 3D eigenfunctions without loosing essential 
information: {\em ray --- angular momentum representation}.
In the momentum --- angular momentum representation, quantum states 
$\Ket{\Psi}$ are labeled by 
eigenvalues of the three commuting observables, 
$\vec{p}^2,(\vec{r}\times\vec{p})^2$, and $(\vec{p}\times\vec{r})_z$,
namely $\hbar^2 k^2,\hbar^2 l(l+1)$, and $\hbar m$,
$$
\Braket{klm}{\Psi} = 
\int\limits_{4\pi}\!d^2\vec{n}\, Y_{lm}(\vec{n})\int\limits_0^{r_B(\vec{n})}
\!dr r^2j_l(kr)\Psi(\vec{r}).
$$
Note that eigenstates in the momentum --- angular momentum representation,
normalized as $\sum_{lm}\int_0^\infty dk k^2 |\Braket{klm}{\Psi^{(n)}}|^2$,
are strongly localized around $k^{(n)}$ in the momentum coordinate $k$ ,
which is classically a constant of the motion.
The structure of eigenfunctions going perpendicularly to the energy shell 
$k\approx k^{(n)}$ is somehow trivial 
($\Braket{klm}{\Psi} \approx 0$, if $|k-k^{(n)}|r_{\text max} \gg 1$)
and certainly less interesting than the structure inside 
the shell. Therefore we define a ray---angular momentum representation of
3D wavefunctions 
\begin{equation}
R_{lm}[\Psi] = \int\limits_0^\infty dk k^2 |\Braket{k l m}{\Psi}|^2, 
\label{eq:ray}
\end{equation}
which gives the total probability for a particle with arbitrary
momentum $\hbar k$ moving along a ray with fixed 
angular momentum ($l,m$ are the natural coordinates on the spherical energy shell
in the momentum space).
Numerical calculations with formula (\ref{eq:ray}) are very tedious since they
require calculation of the full momentum---angular momentum representation
$\Braket{klm}{\Psi}$ first. 
Using well known relations for spherical harmonics and Bessel functions one can
reformulate eq. (\ref{eq:ray}) giving more practical expression
$$
R_{lm}[\Psi] = \frac{\pi}{2}\int\limits_0^{r_{\text max}}\!dr r^2
\left(\int\limits_{4\pi}d^2\vec{n}\, 
\theta(r_B(\vec{n}) - r)Y_{lm}(\vec{n})\Psi(r\vec{n})\right)^2$$
For the eigenstates $\Psi^{(n)}$ of our billiard 
we have $R^{(n)}_{lm} = 0$ unless $l=2n,m=4j$ and
$R^{(n)}_{l,m} = R^{(n)}_{l,-m}$ (due to the symmetry), 
and the integration over $r$ can be simplified up to the radius of the largest 
inscribed sphere,  $r_{\text min} = \min\{r_B(\vec{r}/r)\}$,
\begin{eqnarray*}
R^{(n)}_{l,4j} &=&
\frac{\pi}{2}
\left(\sum\limits_{\alpha}^{l_\alpha=l} \psi^{(n)}_\alpha c_{\alpha j}\right)^2 
\int\limits_0^{r_{\text min}} dr r^2 j^2_l(k^{(n)}r) \\
&+& \frac{\pi}{2}
\int\limits_{r_{\text min}}^{r_{\text max}}\!dr r^2 
\left[\int\limits_{4\pi}d^2\!\vec{n}\, \theta_\epsilon(r_B(\vec{n})-r)
Y_{l,4j}(\vec{n})\Psi^{(n)}(r\vec{n})\right]^2. 
\end{eqnarray*} 
The numerical integrals over $r$ have been performed using another (sufficiently
dense) Gaussian quadrature, and smooth (Fermi) approximation for the step function
$\theta_\epsilon(x) = 1/(1 + \exp(-x/\epsilon))$, where $\epsilon < 2\pi/k^{(n)}$.
For chaotic (microcanonical) eigenfunctions, 
$R^{(n)}_{lm}$ are expected to fluctuate around smooth microcanonical classical
angular momentum distribution 
$R^{(cl)}_{lm} = \ave{\delta(\hbar l - |\vec{r}\times\vec{p}|)
\delta(\hbar m - (\vec{r}\times\vec{p})_z)}$
on the classically allowed region of $lm$ where $R^{(cl)}_{lm}$ is nonzero. 
Strong systematic deviations indicate localization or scarring.
However,
the statistical fluctuations of $\rho_{lm}=R^{(n)}_{lm}/R^{(cl)}_{lm}$ are very large.
We have numerical evidence for the {\em stationarity} and {\em Lorentzian tails} of 
the statistical distribution of $\rho_{lm}$, $d{\cal P}/d\rho(\rho\gg 1) = 
{\cal O}(\rho^{-2})$, 
which has been studied numerically for the full stretch of
64 chaotic eigenfunctions.

In Figure 2 we show the sample of 12 states in the ray---angular momentum 
representation, which is found to be very {\em sensitive to
localization}, 
since the coordinates $l$ and $l_z=m$ are the `most adiabatic' observables, 
(they are the canonical actions for the spherical billiard).
Because of strong fluctuations we show 
(above the diagonals $l=m$ with 
$l$ and $m$ coordinates interchanged) also the smoothed ray-angular momentum 
representation
$$ \tilde{R}^{(n)}_{lm} = \frac{1}{2\pi\Delta^2}\sum_{l'm'} R^{(n)}_{l'm'} 
\exp[-((l-l')^2 + (m-m')^2)/(2\Delta^2)],$$
with $\Delta=3$.
Although the system is classically strongly chaotic, having almost GOE spectral statistics
\cite{P96}, we see lots of significantly localized eigenstates.
Only the states $1,2,3,$ and $9$ are really uniformly extended, the states $7$ and $10$
are localized on the (2+2)D invariant manifold ${\cal T}$, which is mapped 
on the line $m=l/\sqrt{3}$, the states $7,8$ and $10$ are localized on
the symmetric (2+2)D invariant manifolds ${\cal R}_{x,y,z}$, which are mapped on 
the lines $m=0$ and $m=l$. 
This representation is convenient also because classical periodic orbits
are sets of points since each line segment of an orbit has fixed angular momentum.
The state 12 is a strongly enhanced in the neighborhood of an
unstable periodic orbit of desymmetrized billiard
of length $l=10.209$ with a pair of unimodular, $\exp(\pm i 1.558)$, and a pair of real, 
$4.000,0.250,$ eigenvalues of a monodromy matrix.

Brute-force approach: {\em 3D eigenfunctions} in configuration space.
Finally, we have calculated full 3D wavefunctions for the interesting representative
eigenstates of our billiard using expansion (\ref{eq:expansion}) on a dense mesh 
of points in 3D configuration space.
 At every point in space we have then smoothed the probability density
$|\Psi|^2$ by averaging over a ball of radius equal to one de Broglie wavelength $2\pi/k$.
The regions of high smoothed probability density (being above certain threshold) 
have been plotted in 3D for states 1,12, and 8, in figures 3,4, and 5, respectively.
The probability distributions of fluctuation amplitudes $\Psi(\vec{r})$ have been
calculated and compared to a Gaussian \cite{B77}.
The state 1 (figure 3) is shown as an example of a typical chaotic eigenstate 
which is uniformly extended over 
the billiard domain with a Gaussian amplitude distribution $d{\cal P}/d\Psi$.
The state 12 (figure 4) is shown as an example of a 1D scar localized in a 
neighborhood of an unstable periodic orbit.
The state 8 (figure 5) is shown as an example of a 2D scar localized in a neighborhood 
of an (2+2)D invariant chaotic submanifold ${\cal R}_z$. 
(Note that the states $7$ and $10$ have a similar structure but they
are associated with two invariant (2+2)D chaotic submanifolds
simultaneously, ${\cal R}_z$ and ${\cal T}$). Also the amplitude distribution
$d{\cal P}/d\Psi$ for the localized states $8$ and $12$ is different from a Gaussian:
there is a stronger peak for small amplitudes and slowly decaying  tails for 
large amplitudes. 
\\\\
In a conclusion we wish to emphasize the twofold goal of this paper:
(i) the survey of very-high lying eigenstates of a strongly chaotic and generic 3D
billiard which has been made possible by (ii) an application  of the
scaling method for quantization of billiards (which works for the Helmholz equation
in arbitrary dimension), proposed recently \cite{VS95}. 
In a preceding paper \cite{P96} we found small but significant deviations of the
level statistics from the expected ultimate asymptotics (GOE) which have been 
explained in terms of localization of eigenfunctions. Indeed, besides the majority
of extended states with Gaussian random amplitude, we found examles of 1D scars
which are strongly enhanced in the vicinity of short unstable classical periodic orbits,
and examples of 2D scars which are strongly enhanced in the vicinity of the
projections of (2+2)D invariant chaotic submanifolds on to configuration space.
This is the first numerical study of the structure of eigenstates of higher dimensional
($D>2$) chaotic systems in the semiclassical regime (i.e. very large sequential quantum
number or small effective Planck's constant). 
Although the phase space is dynamically more `connected' than in 2D systems,
and the classical dynamics is really uniformly ergodic on a scale of $t_{\text break}$,
we have found very rich behavior ranging from Gaussian pseudo-random
extended states to various types of localized states which are associated with classically 
invariant structures. This is the result which should motivate further theoretical and 
numerical study of quantum chaos in higher dimensional systems.

\section*{Acknowledgments}
Enlightening discussions with H.Primack, E.Vergini and M.Saraceno as
well as financial support of the Ministry of Science and Technology of
the Republic of Slovenia are gratefully acknowledged.

\section*{Figure captions}
\noindent {\bf Figure 1:}
We show the normal derivative of the wavefunctions for the
sample of 12 eigenstates on the mapped
boundary surface, the triangle $0  \le \chi \le \phi < \pi/4$ (see text).
Below the diagonals (abscissa $\phi$, ordinate $\chi$) we show the
magnitude of the normal derivative (using 10 levels of greyness which change when
the square of the function changes for a factor $2.3$),
and above the diagonals (abscissa $\chi$, ordinate $\phi$) we show
the nodal lines of the normal derivative.
The full curves going through the middle of triangles are the images of (2+2)D
invariant submanifold ${\cal T}$, whereas the images of symmetric (2+2)D 
invarinat submanifolds are the boundary lattices $\chi=0,\chi=\phi,$ and $\phi=\pi/4$.
\bigskip
\bigskip

\noindent {\bf Figure 2:}
Same as in figure 2 but for the ray---angular momentum representation of
eigenstates $R^{(n)}_{lm}$ bellow the diagonals 
(abscissa $0\le l\le k^{(n)}r_{\text max}$, ordinate $0 \le m \le m$) and
their smoothed counterparts $\tilde{R}^{(n)}_{lm}$ above
the diagonals (abscissa $m$, ordinate $l$).
The lines $m = l/\sqrt{3}$ are the images of the relevant (2+2)D
invariant manifold ${\cal T}$
and the image of the relevant periodic orbit for the state 12 is
indicated 
with small circles. (See text.)
\bigskip
\bigskip

\noindent {\bf Figure 3:}
The smoothed 3D probability density of a chaotic eigenfunction 
(state 1 on figs. 1 and 2).
The desymmetrized billiard (we plot the curves along its six edges connecting the
four corners $0$ (below), $A_1$ (left), $A_2$ (middle), and $A_3$
(right) as small circles) 
is sliced with horizontal planes $v=(y-z)/\sqrt{2}=n\Delta v,$ for 
$\Delta v = 0.005$ and $n=0,1\ldots 90$, into triangular sections. 
Going from the bottom $n=0$ to the top $n=90$, 
the regions where the smoothed probability density exceeds a 
threshold, $\Psi^2_{\text thr} = 1.1/V$, 
are plotted and filled with the level of greyness
which is proportional to the vertical coordinate $v$ (dark is bottom and bright is top 
--- see grey scale on the right).
Hopefully, one can thus get an impression of the 3D structure of eigenfunctions.
The inset on the lower right corner of the picture shows the histogram of the
amplitude distribution $d{\cal P}/d\Psi$ which is in this case in agreement with 
a Gaussian (dotted curve; vertical dotted lines indicate the positions of $\pm 1 $ 
standard deviation).
\bigskip
\bigskip

\noindent {\bf Figure 4:}
The same as in figure 3 but for a strongly localized --- scarred eigenfunction
(state 12 on figs. 1 and 2). The state is strongly enhanced in the vicinity of
an unstable periodic orbit (see figure 2 and text), so we have picked a higher
threshold $\Psi^2_{\text thr} = 1.55/V$, and the amplitude distribution largely
deviates from a Gaussian.
\bigskip
\bigskip

\noindent {\bf Figure 5:}
The same as in figure 3 but for a strongly localized eigenfunction
(state 8 on figs. 1 and 2) and the same threshold
$\Psi^2_{\text thr} = 1.55/V$ as in fig. 4.
The state is strongly enhanced on the projection of a (2+2)D
invariant chaotic submanifold ${\cal R}_z,\; z = 0,$ so it is a beautifull example of 
a 2D scar.
Correspondingly, the amplitude distribution deviates from a Gaussian.
\bigskip
\bigskip


\begin{thebibliography}{99}
\bibitem[$1$]{BTUPR} {\sc O. Bohigas, S. Tomsovic and D. Ullmo},
Phys. Rep. {\bf 223}, 4 (1993);
{\sc T. Prosen and M. Robnik}, J. Phys. A {\bf 27}, 8059 (1994);
{\sc T. Prosen} Physica D{\bf 91}, 244 (1996).
\bibitem[$2$]{P94} {\sc T. Prosen}, J. Phys. A {\bf 27}, L569 (1994);
{\sc T. Prosen}, Ann. Phys. (N.Y.) {\bf 235}, 115 (1994);
{\sc T. Prosen and M. Robnik}, J. Phys. A {\bf 26}, L319 (1993).
\bibitem[$3$]{SCV} {\sc A. I. Shnirelman}, Usp. Mat. Nauk. {\bf 29}, 181 (1974);
{\sc Y. Colin de Verdiere}, Comm. Math. Phys. {\bf 102}, 497 (1985).
\bibitem[$4$]{B77} {\sc M. V. Berry}, J. Phys. A {\bf 12}, 2083 (1977).
\bibitem[$5$]{H84} {\sc E. J. Heller}, Phys. Rev. Lett. {\bf 53}, 1515 (1984).
\bibitem[$6$]{BBAF} {\sc E. Bogomolny}, Physica {\bf D31}, 169 (1988); 
{\sc M. V. Berry}, Proc. R. Soc. Lond. A {\bf 423}, 219 (1989); 
{\sc O. Agam and S. Fishman},  Phys. Rev. Lett. {\bf 73},806 (1994).
\bibitem[$7$]{PR93} {\sc T. Prosen and M. Robnik}, J. Phys. A {\bf 26}, 5365 (1993).
\bibitem[$8$]{LR94} {\sc Baowen Li and M. Robnik}, J. Phys. A {\bf 27}, 5509 (1994).
\bibitem[$9$]{AOCO} {T. M. Antonsen, E. Ott, Q. Chen, and R. N. Oerter},
Phys. Rev. E {\bf 51}, 111 (1995).
\bibitem[$10$]{VLB47} {\sc F.C. Von der Lage and H.A. Bethe}, Phys. Rev. {\bf 71}, 
612 (1947).
\bibitem[$11$]{P96} {\sc T. Prosen}, 
``Quantization of generic chaotic 3D billiard with smooth boundary I: 
energy level statistics'', {\em Preprint}, submitted to Phys. Lett. A
\bibitem[$12$]{VS95} {\sc E. Vergini and M. Saraceno}, 
Phys. Rev. E {\bf 52}, 2204 (1995).
\bibitem[$13$]{T64} {\sc M. Tinkham}, {\em Group Theory and Quantum Mechanics},
(McGraw-Hill, New York, 1964), p110.
\end{thebibliography}
\end{document}